\documentclass[twocolumn,prl,showpacs,superscriptaddress,preprintnumbers,amsmath,amssymb]{revtex4}
\pdfoutput=1

\usepackage{graphicx}
\begin{document}

\title{Image charge screening: a new approach to enhance magnetic ordering temperatures}

\author{S.~Altieri}
  \affiliation{INFM-CNR-S3, National Center on NanoStructures and BioSystems at Surfaces (S$^3$),
   via G. Campi 213/a, I-41100 Modena, Italy}
\author{M.~Finazzi}
  \affiliation{Dipartimento di Fisica, Politecnico di Milano,
   Piazza Leonardo da Vinci 32, 20133 Milano, Italy}
\author{H.~H.~Hsieh}
  \affiliation{Chung Cheng Institute of Technology, National Defense University,
   Taoyuan 335, Taiwan}
\author{M.~W.~Haverkort}
  \affiliation{II. Physikalisches Institut, Universit{\"a}t zu K{\"o}ln,
   Z{\"u}lpicher Str. 77, D-50937 K{\"o}ln, Germany}
\author{H.-J.~Lin}
  \affiliation{National Synchrotron Radiation Research Center, Hsinchu 30077, Taiwan}
\author{C.~T.~Chen}
  \affiliation{National Synchrotron Radiation Research Center, Hsinchu 30077, Taiwan}
\author{S.~Frabboni}
  \affiliation{INFM-CNR-S3, National Center on NanoStructures and BioSystems at Surfaces (S$^3$),
   via G. Campi 213/a, I-41100 Modena, Italy}
  \affiliation{Department of Physics, University of Modena and Reggio Emilia,
   via G. Campi 213/A, I-41100 Modena, Italy}
\author{G.~C.~Gazzadi}
  \affiliation{INFM-CNR-S3, National Center on NanoStructures and BioSystems at Surfaces (S$^3$),
   via G. Campi 213/a, I-41100 Modena, Italy}
\author{A.~Rota}
  \affiliation{Department of Physics, University of Modena and Reggio Emilia,
   via G. Campi 213/A, I-41100 Modena, Italy}
\author{S.~Valeri}
  \affiliation{INFM-CNR-S3, National Center on NanoStructures and BioSystems at Surfaces (S$^3$),
   via G. Campi 213/a, I-41100 Modena, Italy}
  \affiliation{Department of Physics, University of Modena and Reggio Emilia,
   via G. Campi 213/A, I-41100 Modena, Italy}
\author{L.~H.~Tjeng}
 \affiliation{II. Physikalisches Institut, Universit{\"a}t zu K{\"o}ln,
  Z{\"u}lpicher Str. 77, D-50937 K{\"o}ln, Germany}

\date{\today}

\begin{abstract}

We have tested the concept of image charge screening as a new
approach to enhance magnetic ordering temperatures and superexchange
interactions in ultra thin films. Using a 3 monolayer NiO(100) film
grown on Ag(100) and an identically thin film on MgO(100) as model
systems, we observed that the N\'{e}el temperature of the NiO film
on the highly polarizable metal substrate is 390 K while that of the
film on the poorly polarizable insulator substrate is below 40 K.
This demonstrates that screening by highly polarizable media may
point to a practical way towards designing strongly correlated oxide
nanostructures with greatly improved magnetic properties.

\end{abstract}


\maketitle

Transition metal oxides exhibit many spectacular magnetic and
electrical properties including high temperature superconductivity
and colossal magnetoresistance \cite{Iamada98} making them
particularly promising for nanoscience technology applications. An
acute issue in the field of nanoscience, however, is the strong
reduction of the relevant critical or ordering temperatures due to
well known finite size effects
\cite{Abarra96,Punnoose01,Zheng04,Alders98}. If ways could be found
to compensate for these reductions, one would immediately enlarge
the materials basis for nano-technology. Current approaches to
overcome these problems include the use of chemical doping,
pressure, and strain
\cite{Bloch66,Zhang06,Sidorov98,Massey90,Kaneko87,Locquet98}.

Here we propose to exploit image charge screening as a new method to
compensate finite size phenomena and to enhance magnetic ordering
temperatures well beyond the capability of conventional methods
\cite{Bloch66,Zhang06,Sidorov98,Massey90,Kaneko87,Locquet98}. The
basic idea is to bring the material in the close proximity of a
strongly polarizable medium. The relevant exchange and superexchange
interactions, and thus the related magnetic ordering temperatures,
can then be substantially amplified by reducing the energies of the
underlying virtual charge excitations as a result of the
image-charge-like screening by the polarizable medium
\cite{Hesper97,Altieri99,Duffy83}.

To prove this concept we have chosen to measure the N\'{e}el
temperature $T_N$ of a 3 monolayer (ML) NiO film epitaxially grown
on a MgO(100) substrate and of an equally thin film on Ag(100). NiO
on MgO and on Ag are ideal model systems for this study because of
their simple crystal structure and well characterized growth
properties. They have a rock-salt crystal structure with lattice
constant $a_{MgO}$ = 4.212~\AA~and $a_{NiO}$ = 4.176 \AA,
respectively, corresponding to a lattice misfit of about 1\%. This
ensures a perfect layer-by-layer epitaxial growth of NiO(100) on
MgO(100), with a NiO(100) film surface roughness of about
0.1~\AA~\cite{James99}. Silver has a cubic {\it fcc} structure with
a lattice constant $a_{Ag}$ = 4.086 \AA~and a mismatch with respect
to NiO of about 2\%. When misfit dislocations are avoided by keeping
the film thickness below the critical thickness for strain
relaxation (about 30 ML for NiO/Ag \cite{Giovanardi03}) as done in
the present work, then NiO(100) films grow on Ag(100) in a nicely
layered and coherent mode with a sharp interface. This was already
demonstrated by Kado \cite{Kado95,Kado94}, but it has also been
verified on our samples. We find that the NiO/MgO system has $T_N$
$<$ 40 K, indeed very much reduced from the $T_N$ = 523 K bulk
value. For the NiO/Ag sample, by contrast, we find $T_N$ = 390 K,
showing that the dramatic finite-size effects can be almost
counterbalanced by the near presence of the metal.

The NiO/MgO and NiO/Ag samples were prepared and comparatively
studied {\it in situ} under identical conditions using the Cologne
University MBE-XAS set up at the Dragon beam line of the NSRRC in
Taiwan. Microscopy experiments were performed at the Modena
University. Stoichiometric NiO films were grown by
atomic-oxygen-assisted reactive deposition on a highly ordered
Ag(100) and on cleaved MgO(100) single crystals kept at 463 K in a
background oxygen pressure of 5$\cdot$10$^{-7}$~mbar (base pressure
~3$\cdot$10$^{-10}$~mbar). Immediately after the film growth, both
the NiO/MgO and the NiO/Ag samples where capped {\it in situ} with a
protective 25 ML thick MgO overlayer. The thickness of both the NiO
film and the MgO capping layer was calibrated by monitoring the
intensity oscillations of reflection high energy electron
diffraction (RHEED) in preceding NiO and MgO film growth
experiments. Ex-situ high resolution transmission electron
microscopy (HRTEM) measurements provided an independent absolute
calibration of the NiO and MgO film thickness and a detailed
structural characterization of 500 nm Ag(100)/2.3 nm
NiO(100)/Ag(100) by studying cross sectional lamellae 80 nm thick
obtained by focussed ion beam milling (FEI Strata 235DB) and
lift-out extraction (Kleindeik MM3A micromanipulators).

\begin{figure}
\includegraphics[width=0.45\textwidth]{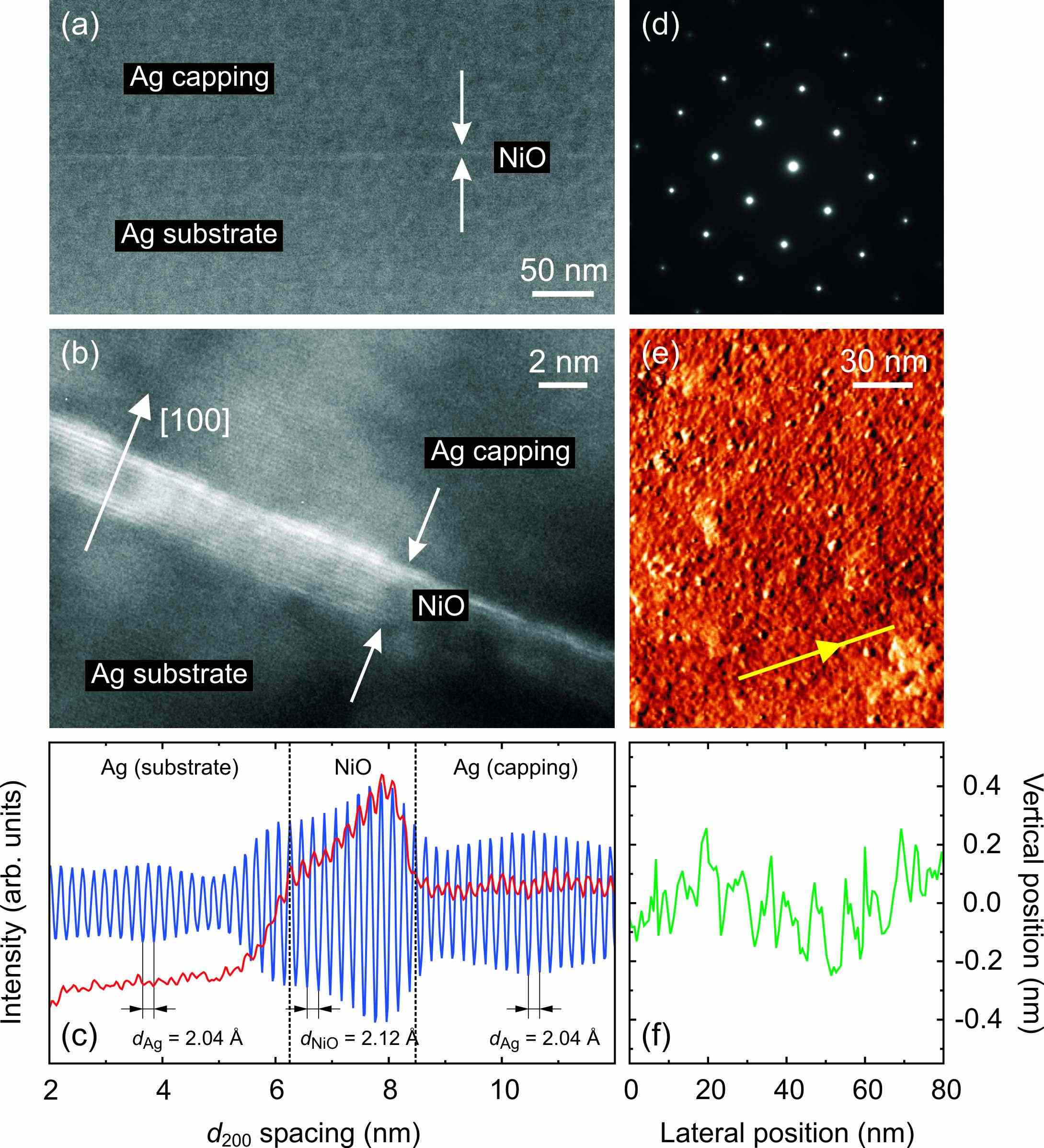}
\caption{Ag/NiO/Ag cross sectional lamella: (a) SEM image; (b)
HRTEM image; (c) HRTEM intensity line scan through Ag/NiO/Ag
interfaces; (d) SAD pattern. (e) STM image of uncapped 3 ML
NiO(100)/Ag(100) at $V$=3 V and $I$=0.1 nA. (f): STM intensity
line scan along the yellow line in panel (e)} \label{f1}
\end{figure}

Fig. 1(a) displays a cross sectional scanning electron microscopy
(SEM) image of a Ag/NiO/Ag sandwich recorded at micrometric length
scale showing that the NiO film covers uniformly the Ag(100) single
crystal. Fig. 1(b) displays a HRTEM image of the Ag/NiO/Ag sandwich
where the atomic structure due to the (200) planes along the [100]
direction is clearly resolved. Intensity line scans across the
Ag/NiO/Ag interfaces are reported in Fig. 1(c). The red and blue
line were measured on the raw data and on the Bragg filtered image,
respectively, and consistently give the same result: the fringes
measured both in the Ag substrate and capping layer have identical
spacing ($d_{Ag}$ = 2.04 \AA) while those measured within the NiO
film region have a larger average spacing ($d_{NiO}$ = 2.12 \AA) by
an amount $\varepsilon^{exp}_{\bot}$ = 3.9\%. The  theoretically
expected strain state of fully coherent NiO/Ag epitaxial film with
tetragonal distortion is given by $\varepsilon^{th}_{\bot}$ = $f$
(1+$\nu$) / (1-$\nu$), where $f$ = ($a_{NiO}$-$a_{Ag}$)/$a_{Ag}$ and
$\nu$ is the NiO Poisson ratio \cite{Hornstra78}. Reported values of
$\nu$ are between 0.21 \cite{James99} and 0.30 \cite{Giovanardi03}
which yield $\varepsilon^{th}_{\bot}$ between 3.4\% and 4.1\%, in
good agreement with the measured value. The intensity profile in
Fig. 1(c) also shows that the transition from Ag to NiO occurs over
a length scale of about one lattice spacing. Fig. 1(d) displays a
selected area diffraction (SAD) pattern recorded with the HRTEM
electron beam impinging over an area of about 1~$\mu$m centred on
the NiO layer, showing the highly ordered crystal structure of the
entire Ag/NiO/Ag sandwich. Fig. 1(e) displays a scanning tunnelling
microscopy (STM) image of an uncapped 3 ML NiO/Ag film showing that
the NiO layer covers uniformly flat silver terraces with an average
surface roughness of 1.8 $\pm$ 0.2 \AA. There are no signs in Fig.
1(e) of three-dimensional NiO islanding on the Ag substrate, in
agreement with the layered nature revealed by the SEM and HRTEM data
in Fig. 1(a-c).

\begin{figure}
\includegraphics[width=0.3\textwidth]{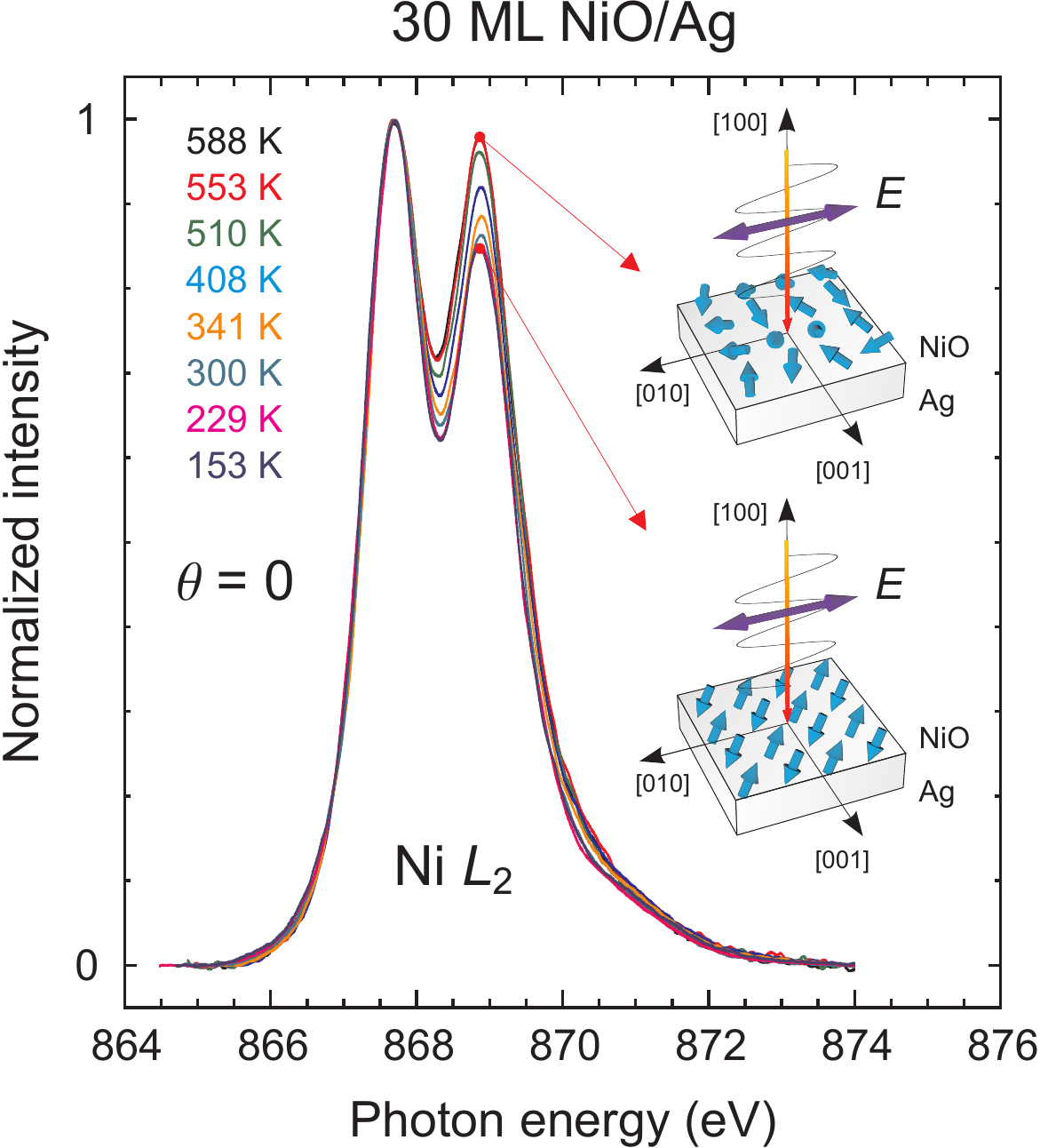}
\caption{Temperature dependent Ni $L_2$ XAS spectra with linearly
polarized light at $\theta$=0$^{\circ}$ on 20 ML MgO(100)/30 ML
NiO(100)/Ag(100) and related changes in magnetic ordering.}
\label{f2}
\end{figure}

To investigate the magnetic properties of these NiO thin films, we
have utilized the magnetic linear dichroic (MLD) effect in the Ni
$L_2$ x-ray absorption spectra (XAS). Experimental and theoretical
studies in the last decade
\cite{Alders98,Alders95,Kuiper93,Ohldag01,Altieri03,Csiszar05,Arenholz07}
have demonstrated that MLD-XAS is a very powerful method to
investigate thin film magnetic properties, especially as far as
antiferromagnetic materials are concerned. To illustrate the
application of this technique specifically for our case, we start
with the XAS spectra for a 30 ML NiO/Ag film. Fig. 2 displays the
temperature dependence of the spectra recorded with linearly
polarized light with the Poynting vector making an angle
$\theta$=0$^{\circ}$ with the sample surface normal, and with the
electric field E lying within the (010) plane. The intensity of
the peak at 868.9 eV photon energy relative to that at 867.7 eV
changes with temperature by about 15-20\%. Although quite small,
this effect is however highly reproducible and is at least an
order of magnitude larger than the noise. The effect can thus be
trusted and, following earlier work on NiO thin films on MgO
\cite{Alders95,Alders98} can be associated with the occurrence of
antiferromagnetic order. By plotting the intensity ratio between
the two peaks at 867.7 and 868.9 eV ($L_2$ ratio) as a function
of temperature, as done in the top panel of Fig. 3, one obtains a
direct measure of the long range order parameter and the N\'{e}el
temperature of the material. For the 30 ML NiO/Ag film we thus
measure $T_N$ = 535 K, which is close to the bulk value of 523 K
\cite{Slack60}. Apparently, the 30 ML NiO/Ag film is already thick
enough to behave as the bulk oxide and not to feel any longer the
influence of the underlying Ag substrate \cite{Altieri99}.

Fig. 4(a) displays the temperature dependent XAS spectra of the 3
ML NiO/Ag film with the Poynting vector making an angle
$\theta$=75$^{\circ}$ with the sample surface normal. The
intensity of the peak at 868.9 eV relative to that at 867.7 eV
clearly changes with temperature. It is important to stress that
the spectra are highly reproducible also for this ultra thin NiO
film. This is shown in Fig. 4(b) reporting the first spectrum
measured at 298 K just after the film growth and the last
spectrum measured at 293 K after six thermal cycles covering the
range from 138 K to 488 K. In the same panel we also show the
spectra measured at $T$ = 473 K and $T$ = 143 K in the first
thermal cycle, and the corresponding spectra measured at $T$ = 488
K and $T$ = 138 K in the last thermal cycle proving the excellent
thermal and chemical stability of the studied samples.  The
effect in Fig. 4(a) thus reflects the temperature dependence of
the antiferromagnetic order. The bottom panel of Fig. 3 reports
the $L_2$ ratio as a function of temperature, and it reveals that
the N\'{e}el temperature of the 3 ML NiO/Ag film is $T_N$ = 390
K, which is about 75\% of the bulk value. As we will show below,
it is surprising that the reduction of the N\'{e}el temperature in
such a thin NiO film is rather modest.

\begin{figure}
\includegraphics[width=0.3\textwidth]{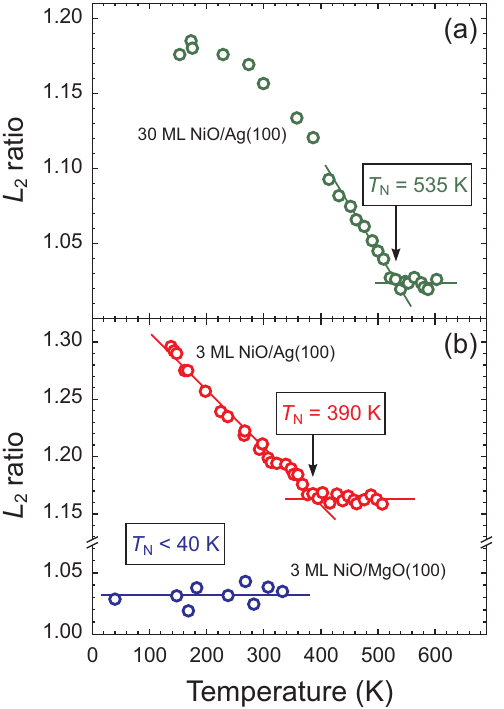}
\caption{Temperature dependent Ni $L_2$ ratio of (a) 30 ML
NiO(100)/Ag(100) and (b) 3 ML NiO(100)/Ag(100) (red markers) and 3
ML NiO(100)/MgO(100) (blue markers).} \label{f3}
\end{figure}

\begin{figure}
\includegraphics[width=0.4\textwidth]{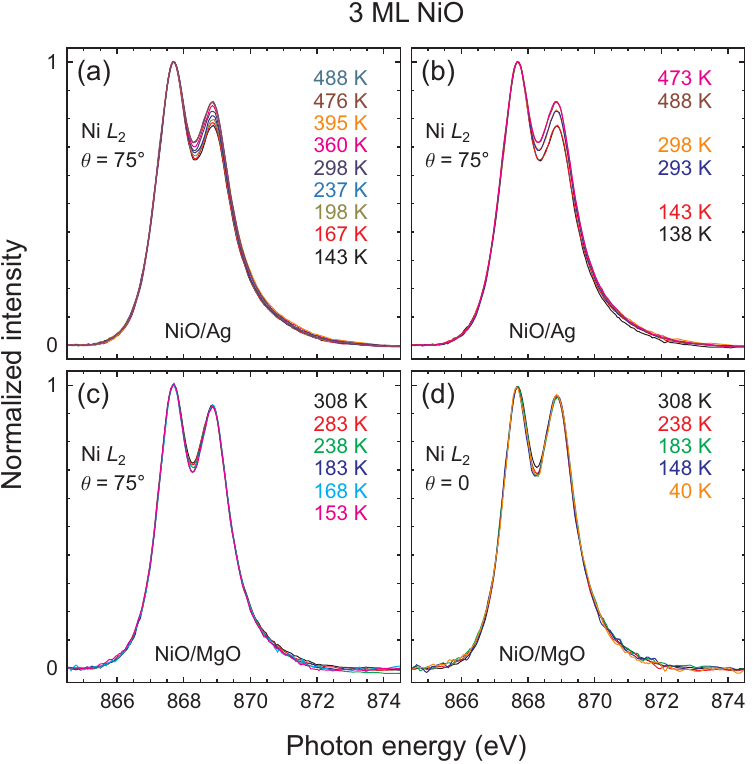}
\caption{Temperature dependent Ni $L_2$ XAS spectra on (a)-(b)  25
ML MgO(100)/3 ML NiO(100)/Ag(100) and (c-d) 25 ML MgO(100)/3 ML
NiO(100)/MgO(100) films.} \label{f4}
\end{figure}

Figs. 4(c-d) display the spectra of the 3 ML NiO/MgO film for
temperatures between 40 K and 308 K with $\theta$=75$^{\circ}$
[Fig. 4(c)] and $\theta$=0$^{\circ}$ [Fig. 4(d)]. Apart from a
minor broadening effect with temperature, not much else is
actually happening with the spectra. In particular, the intensity
ratio between the two peaks at 867.7 eV and 868.9 eV is
essentially temperature independent. In comparing the
$\theta$=75$^{\circ}$ with the $\theta$=0$^{\circ}$ spectra, we
also notice that the intensity ratio between the two peaks does
not change, in very strong contrast with the case for 20 ML
NiO/MgO \cite{Alders95,Alders98}. All this suggests that there is
no magnetic order in the 3 ML NiO/MgO, at least not in the
measured temperature range. This, in turn, means that the N\'{e}el
temperature $T_N$ is dramatically suppressed, from 523 K for bulk
NiO \cite{Slack60} to well below 40 K for 3 ML NiO/MgO. This
finite-size effect in our NiO/MgO film is not inconsistent with an
earlier thickness dependent study finding T$_N$ = 470 K, 430 K,
and 295 K for 20 ML, 10 ML, and 5 ML NiO/MgO films, respectively
\cite{Alders98}.

Although the N\'{e}el temperature of 3 ML NiO/Ag (390 K) is lower
than that of bulk NiO (523 K), it is still at least ten times
larger than that of 3 ML NiO/MgO ($<$ 40 K), see Fig. 3(b). The
dramatic finite size effects in these ultra thin films are
manifest in NiO/MgO, but are efficiently counter balanced in the
NiO/Ag by another effect that can only be related to the presence
of the Ag substrate.

To discuss the origin of this extraordinary large enhancement
factor, we have to consider how the superexchange interactions in
the NiO films are modified by the presence of the substrate.
Using the well known extended Anderson expression for the
superexchange coupling constant $J$
\cite{Anderson87,Zaanen85,Zaanen87}:

\begin {equation}
J = [(-2t^4/\Delta^2)] [(1/\Delta)+(1/U)]
\end {equation}

\noindent in which $t$ is the anion $2p$ - cation $3d$ transfer
integral, $\Delta$ the $2p-3d$ charge transfer energy, and $U$ the
on-site $3d$ Coulomb energy, one can expect that $J$, and
consequently $T_N$, can be amplified by increasing $t$ and/or
decreasing $\Delta$ or $U$.

Already two decades ago, Duffy and Stoneham \cite{Duffy83}
predicted that a medium with a high dielectric polarizability
should provide an effective screening for various charge
excitations in a nearby located material. Indeed, Hesper et al.
\cite{Hesper97} and Altieri et al. \cite{Altieri99} have shown
spectroscopically that the band gap, the Hubbard-$U$, and the
charge-transfer-$\Delta$ energies of an insulating material can
be strongly reduced from its bulk values by depositing it as a
thin film on top of a metal substrate. It has been estimated that
this so-called image charge screening effect could result in a
50\% reduction of $U$ and $\Delta$ for NiO on Ag
\cite{Altieri99}. For NiO on MgO one can even envision an opposite
effect: the polarizability of MgO is less than that of NiO, with
the result that the $U$ and $\Delta$ parameters in the NiO films
are increased in comparison to the bulk values. Based on these
large screening effects, it becomes almost obvious to expect a
substantial enhancement of $J$ and $T_N$ in NiO/Ag compared to
NiO/MgO, especially considering that $U$ and $\Delta$ parameters
enter as 1/$\Delta^2$$U$ and 1/$\Delta^3$ into equation (1).

We notice that the use of hydrostatic pressure to reduce
isotropically the interatomic spacing and thus to increase $t$ is
the conventional method to enhance superexchange interactions and
the N\'{e}el temperature in many transition metal oxides such as
NiO, CoO, FeO, and MnO
\cite{Bloch66,Zhang06,Sidorov98,Massey90,Kaneko87}, as well as to
increase $T_c$'s in various high temperature superconductors
\cite{Kaneko87}. Yet, the influence of modified $t$ on $J$ and
$T_N$ in our NiO/Ag and NiO/MgO films is negligible as we will
show now using the theoretical \cite{Zhang06} and experimental
\cite{Sidorov98,Massey90} lattice dependent $J$ and $T_N$ values
in NiO. For bulk NiO with $a_{NiO}$ = 4.10 \AA~ one obtains $J$ =
19.0 meV and $T_N$ = 523 K. If NiO is isotropically forced to fit
the lattice constant of MgO or Ag, so that $a_{NiO}$ = $a_{MgO}$
= 4.21 \AA~or $a_{NiO}$ = $a_{Ag}$ = 4.09 \AA, then $J^{MgO}$ =
18.42 meV and $T_N$$^{MgO}$ = 507 K, or $J^{Ag}$ = 23.2 meV and
$T_N$$^{Ag}$ = 639 K, respectively
\cite{Zhang06,Sidorov98,Massey90,Kaneko87}. This would constitute
an enhancement of both $J$ and $T_N$ by about a factor of 1.3
when comparing NiO/Ag with NiO/MgO, which is by far not enough to
explain the contrast between the values $T_N$ = 390 K and $T_N$
$<$ 40 K measured on NiO/Ag and NiO/MgO, respectively. Moreover,
the strain in our films is non-isotropic, and one could expect
that the lattice spacing effect will be smaller since the change
in the interatomic spacing along the surface normal is opposite
to that in plane. Indeed, it has been experimentally shown that
for NiO and CoO films on MgO uniaxial strain up to 2\% has a
negligible effect on $T_N$ \cite{Abarra96}.

Important for the modelling of our results are the findings of
recent \textit{ab-initio} density-functional band structure
calculations on ultra thin NiO films, both free standing and
supported on Ag(100) \cite{Casassa02,Cinquini06}. The calculations
predict that the magnetization and the superexchange constant $J$ of
3 ML NiO films should hardly be affected by a nearby substrate. This
shows that indeed $T_N$ enhancement mechanisms based on an increase
of $t$ via the substrate should not play a major role. Moreover,
this means that the explanation for the experimentally observed
large discrepancy in $T_N$ values for the NiO/Ag and NiO/MgO systems
requires models which go beyond static mean field theories. The
screening model proposed here fulfills exactly this requirement: it
explains the large differences in $T_N$ and it is not captured by
the standard band structure calculations.

To conclude, using NiO/Ag and NiO/MgO as model systems, we have been
able to show that screening by a metallic substrate provides a novel
approach to enhance magnetic ordering temperatures and superexchange
interactions in ultra thin films well beyond the capability of
conventional methods. We note that image charge screening is not
restricted to metals but is also present when using small-gap
semiconductors with high dielectric polarizability. These results
may point to a practical way towards designing strongly correlated
oxide nanostructures with greatly improved N\'{e}el or Curie
temperatures for various forms of magnetic ordering.

We would like to thank G.A. Sawatzky for stimulating this work in
several occasions. SA acknowledges CNR funding through project
Short Term Mobility, and financial support by CNR-INFM- S3
National Centre through project {\it Seed Activity}. The research
in K{\"o}ln is supported by the Deutsche Forschungsgemeinschaft
through SFB 608.


\begin{thebibliography}{10}

\bibitem{Iamada98} M. Imada, A. Fujimori, and Y. Tokura,
   Rev. Mod. Phys. {\bf 70}, 1039  (1998);

\bibitem{Abarra96} E.N. Abarra, K. Takano, F. Hellman, and A.E. Berkowitz,
   Phys. Rev. Lett. {\bf 77}, 3451  (1996);

\bibitem{Punnoose01} A. Punnoose {\it et al.},
   Phys. Rev. B {\bf 64}, 174420 (2001);

\bibitem{Zheng04} X.G. Zheng {\it et al.},
   Solid State Comm. {\bf 132}, 493  (2004);

\bibitem{Alders98} D. Alders, L.H. Tjeng {\it et al.},
   Phys. Rev. B {\bf 57}, 11623  (1998);

\bibitem{Bloch66} D. Bloch,
   J. Phys. Chem. Sol.  {\bf 27}, 881  (1966);

\bibitem{Zhang06} W.-B. Zhang, Y.-L. Hu, K.-L. Han, and B.-Y. Tang,
   Phys. Rev. B {\bf 74}, 54421  (2006);

\bibitem{Sidorov98} V.A. Sidorov,
   Appl. Phys. Lett. {\bf 72}, 2174 (1998);

\bibitem{Massey90} M.J. Massey, N.H. Chen, J.W. Allen, and R. Merlin,
   Phys. Rev. B {\bf 42}, 8776  (1990);

\bibitem{Kaneko87} T. Kaneko, H. Yoshida, S. Abe, H. Morita, K. Noto, and H. Fujimori,
   Jap. J. Appl. Phys.  {\bf 26}, L1374  (1987);

\bibitem{Locquet98} J.-P. Locquet, J. Perret, J. Fompeyrine, E. M{\"a}chler, J. W. Seo, and G. Van Tendeloo,
   Nature  {\bf 394}, 453 (1998)

\bibitem{Hesper97} R. Hesper, L.H. Tjeng, and G.A. Sawatzky,
   Europhys. Lett.   {\bf 40}, 177  (1997);

\bibitem{Altieri99} S. Altieri and L. H. Tjeng, F. C. Voogt and T. Hibma, and G. A. Sawatzky,
   Phys. Rev. B {\bf 59}, R2517  (1999);

\bibitem{Duffy83} D.M. Duffy and A.M. Stoneham
   J. Phys. C  {\bf 16}, 4087  (1983);

\bibitem{James99} M.A. James and T. Hibma,
   Surf. Sci.  {\bf 433-435}, 718 (1999);

\bibitem{Giovanardi03} C. Giovanardi, A. di Bona, S. Altieri {\it et al.},
  Thin Solid Films  {\bf 428}, 195 (2003);

\bibitem{Kado95} T. Kado,
   J. Appl. Phys. {\bf 78}, 3149  (1995);

\bibitem{Kado94} T. Kado,
   J. Crystal Growth  {\bf 144}, 329 (1994);

\bibitem{Hornstra78} J. Hornstra {\it et al.},
   J. Crystal Growth  {\bf 44}, 513 (1978);

\bibitem{Alders95} D. Alders {\it et al.},
   Europhys. Lett. {\bf 32}, 259 (1995).

\bibitem{Kuiper93} P. Kuiper, B.G. Searle, P. Rudolf, L.H. Tjeng, and C.T. Chen,
   Phys. Rev. Lett. {\bf 70}, 1549  (1993);

\bibitem{Ohldag01} H. Ohldag, T.J. Regan, J. St{\"o}hr, A. Scholl, F. Nolting {\it et al.}
   Phys. Rev. Lett. {\bf 87}, 247201  (2001);

\bibitem{Altieri03} S. Altieri, M. Finazzi, H.H. Hsieh {\it et al.}
   Phys. Rev. Lett. {\bf 91}, 137201  (2003);

\bibitem{Csiszar05} S.I. Csiszar, M.W. Haverkort, Z. Hu {\it et al.}
   Phys. Rev. Lett. {\bf 95}, 187205  (2005);

\bibitem{Arenholz07} E. Arenholz, G. van der Laan, R.V. Chopdekar, and Y. Suzuki,
   Phys. Rev. Lett. {\bf 98}, 197201  (2007);

\bibitem{Slack60} G.A. Slack,
  J. Appl. Phys. {\bf 31}, 1571 (1960);

\bibitem{Anderson87} P.W. Anderson,
   Science  {\bf 235}, 1196 (1987); ibid. Phys. Rev. {\bf 115}, 2 (1959)

\bibitem{Zaanen85} J. Zaanen, G.A. Sawatzky, and J.W. Allen,
   Phys. Rev. Lett. {\bf 55}, 418  (1985);

\bibitem{Zaanen87} J. Zaanen and G.A. Sawatzky,
   Can. J. Phys. {\bf 65}, 1262  (1987);

\bibitem{Casassa02} S. Casassa,  A. M. Ferrari, M. Busso, and C. Pisani,
   J. Phys. Chem. B   {\bf 106}, 12978  (2002);

\bibitem{Cinquini06} F. Cinquini, L. Giordano, G. Pacchioni, A.M.
Fer rari, C. Pisani, and C. Roetti
   Phys. Rev. B   {\bf 74}, 165403   (2006);

\end{thebibliography}
\end{document}